\documentclass{article}
\usepackage{ijcai19}

\usepackage{times}
\usepackage{soul}
\usepackage{url}
\usepackage[hidelinks]{hyperref}
\usepackage[utf8]{inputenc}
\usepackage[small]{caption}
\usepackage{graphicx}
\usepackage{amsmath}
\usepackage{booktabs}
\usepackage{algorithm}
\usepackage{algorithmic}
\usepackage{mathtools}
\usepackage{amsthm}
\usepackage{array}
\usepackage{amsfonts}
\usepackage{color}
\usepackage{amssymb}
\usepackage{diagbox}
\usepackage{relsize}
\usepackage{soul}
\usepackage{bm}
\usepackage{multirow, makecell}

\newtheorem{definition}{Definition} 
\newtheorem{theorem}{Theorem}
\newcommand{\citet}[1]{\citeauthor{#1}~\shortcite{#1}}
\newcommand{\citep}{\cite}

\title{SPINE: Structural Identity Preserved Inductive Network Embedding}

\author{
Junliang Guo \and Linli Xu\footnote{Corresponding Author} \And Jingchang Liu \\
\affiliations
Anhui Province Key Laboratory of Big Data Analysis and Application,\\
School of Computer Science and Technology,
University of Science and Technology of China \\
\emails
guojunll@mail.ustc.edu.cn, linlixu@ustc.edu.cn, xdjcl@mail.ustc.edu.cn
}

\begin{document}

\maketitle

\begin{abstract}
Recent advances in the field of network embedding have shown that
low-dimensional network representation is playing
a critical role in network analysis. Most existing network embedding
methods encode the local proximity of a node,
such as the first- and second-order proximities.
While being efficient, these methods are short of leveraging the global structural information
between nodes distant from each other.
In addition, most existing methods learn embeddings on one single fixed network,
and thus cannot be generalized to unseen nodes or networks without retraining.
In this paper we present SPINE, a method that can jointly capture the local proximity and
proximities at any distance, while being inductive to efficiently deal with unseen nodes or networks.
Extensive experimental results on benchmark datasets
demonstrate the superiority of the proposed framework
over the state of the art.
\end{abstract}

\section{Introduction}
\label{sec:intro}

Network embedding has been successfully applied in a wide variety of network-based machine learning tasks, such as node classification, link prediction, and
community detection, etc~\citep{cai2017comprehensive,kipf2016semi}. Different to the primitive network representation, which suffers from overwhelming high
dimensionality and sparsity, network embedding aims to learn low-dimensional continuous latent representations of nodes on a network while preserving the structure and the inherent properties of the network, which can then be exploited effectively in downstream tasks. 

Most existing network embedding methods approximate local proximities via
random walks or specific objective functions,
followed by various machine learning algorithms with specific objective functions
to learn embeddings~\citep{perozzi2014deepwalk,grover2016node2vec}. 
Specifically, the local proximity of a node is approximated with the routine of learning the embedding vector of a node by predicting its neighborhood, inspired by the word embedding principle~\citep{mikolov2013efficient,mikolov2013distributed} which learns the embedding vector of a word by predicting its context.

\begin{figure}[tb]
  \centerline{\includegraphics[width=0.75\columnwidth]{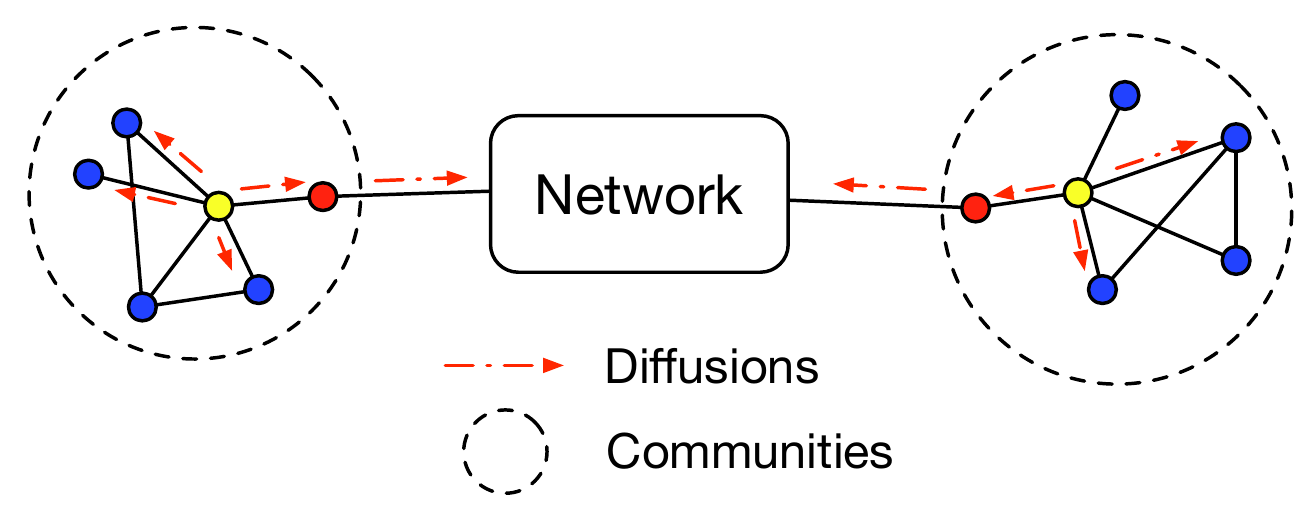}}
  \caption{An example of structural identities in the information diffusion process.
    Red arrows
    indicate information diffusions between nodes (e.g., retweeting in Twitter),
    and nodes in the same circle are in the same community.
  }
  \label{fig:diff}
  \vspace{-10pt}
\end{figure}

However, there still exist some potential issues that need further concerns.
On the one hand, local proximity preserved methods generally do not model nodes
far from each other in practice.
Meanwhile, in real-world network mining tasks,
nodes that are far apart but 
close {\tiny }\emph{in structural identity},
or in other words, take similar \emph{roles}, should perform similarly on specific tasks.
Figure~\ref{fig:diff} shows an example of
how nodes with different roles perform in the information diffusion processes on social networks.
Nodes with different colors indicate different roles in social networks,
i.e., \emph{structural hole spanners}~(red nodes), \emph{opinion leaders}~(yellow nodes)
and \emph{ordinary users}~(blue nodes) respectively~\citep{lou2013mining,yang2015rain}.
Intuitively, nodes with same roles
behave similarly even with a large distance (yellow nodes in Figure~\ref{fig:diff}),
which is the property that should be preserved in the embedding space.
In the meantime, the local proximity of a node is also crucial in network embedding.
For example in Figure~\ref{fig:diff}, nodes in the same community should
be clustered tightly in the embedding space.
Therefore, a desirable network embedding method should preserve the local proximity
and the global structural identity of a node simultaneously to
represent the node precisely.
Unfortunately, most existing methods fail to consider the local and global structural information at the same time. In principle, it is challenging to interactively integrate the two kinds of information to obtain comprehensive embeddings rather than a trivial linear combination.

On the other hand, most existing network embedding approaches are \emph{transductive}.
To be specific, embeddings are learned on a fixed network, and cannot be directly applied
to new joined nodes or other networks. In contrast, \emph{inductive}
methods which are able to generalize to unseen nodes or totally new networks are extensively
required in real-world applications, e.g., social recommendation for new users,
classification of protein functions in various protein-protein interaction graphs, etc.
Unfortunately, traditional network embedding principles such as
random walk and matrix factorization are impracticable for unseen nodes or networks,
which makes inductive network embedding much more challenging than transductive problems.

In this paper, we propose \textbf{SPINE}, an inductive network embedding
framework which jointly preserves local proximities and structural identities of nodes.
We show that structural
similarities between node pairs can be represented by a high-order
proximity of the network known as Rooted PageRank (RPR)~\citep{liben2007link},
and by assigning each node a structural feature vector based on RPR,
we can encode structural proximities between
nodes by measuring the similarities of their structural features.
To construct an inductive framework, we learn an embedding generator rather than directly optimizing a unique embedding for each node,
through which local proximities are integrated.
To further encode structural identities, we propose
a biased Skip-Gram Negative Sampling (SGNS) approach with a novel positive sampling strategy guided by structural
similarities between nodes.
Furthermore,
the objective function of SPINE is carefully designed to enhance the
structural information contained in the embedding generator.

\section{SPINE}
\label{sec:model}

In this section, we propose Structural Identity Preserved Inductive Network Embedding (SPINE), a novel inductive approach for \emph{unsupervised} network embedding,
which consists of three components: structural feature generation, embedding generation
and biased SGNS optimization.

\noindent \textbf{Problem Definition.}~~~
Given an undirected network $G=\{V,E,\bm{F}\}$, in which a set of nodes $V$ are connected
by a set of edges $E$, and $\bm{F} \in \mathbb{R}^{|V| \times f}$ is the content matrix of nodes. 
The adjacency matrix is $\bm{A}$ where
$A_{i,j}=w_{i,j}$ is the edge weight between node $v_{i}$ and $v_{j}$,
and we denote the corresponding transition matrix as $\bm{P}$, where
$P_{i,j} = \frac{w_{i,j}}{\sum_{k=1}^{|V|} w_{i,k}}$ represents the transition
probability between node $v_{i}$ and $v_{j}$.
Our goal is to learn $\bm{E} \in \mathbb{R}^{|V| \times d}$, where $d$ is a small number of latent dimensions.
These low-dimensional representations should well preserve the structural properties of $G$,
including local proximities and structural identities,
which can be evaluated with downstream tasks such as node classification.

\subsection{Rooted PageRank and Structural Identity}
\label{sec:prove}
We start with theoretical preliminaries of our
structural feature generation algorithm.
Here we introduce a well-known high-order
proximity of a network named Rooted PageRank (RPR)~\citep{liben2007link}, defined as
$\bm{S}^{\textrm{RPR}} = (1-\beta_{\textrm{RPR}})(\bm{I}-\beta_{\textrm{RPR}}\bm{P})^{-1}$,
where $\beta_{\textrm{RPR}}\in (0,1)$ is
the probability of the current node randomly walking to a neighbor rather than jumping back
to the start node.
The $(i,j)$-th entry of
$\bm{S}^{\textrm{RPR}}$ is the probability that a random walk from node $v_{i}$ will stop at $v_{j}$
in the steady state, which can be used as an indicator of the node-to-node proximity.
Therefore, we can use the $i$-th row of $\bm{S}^{\textrm{RPR}}$, denoted as $\bm{S}^{\textrm{RPR}}_{i}$,
to represent the global structural information of node $v_{i}$.
We can further rewrite $\bm{S}^{\textrm{RPR}}_{i}$ in a recursive manner as:
\begin{equation} \small
\label{def:RPR}
S_{i}^{\textrm{RPR}} = \beta_{\textrm{RPR}}\bm{P}S_{i}^{\textrm{RPR}} + (1-\beta_{\textrm{RPR}})\bm{g}_{i}
\end{equation}
where $\bm{g}_{i}$ is the index vector of node $v_{i}$ whose $i$-th element is $1$ while others are $0$.

Next, we are going to verify
that $\bm{S}^{\textrm{RPR}}_{i}$ is able to represent the structural identity of node $v_{i}$.
We first define the \emph{complete structural property}~\citep{batagelj1992direct}
of a node as:
\begin{definition}
\label{def:node_proper}
A node property $t \colon V \to \mathbb{R}$ is complete structural if for any automorphism 
$\varphi$
of every node $v_{i} \in V$, it always satisfies:
\begin{displaymath}
t(v_{i}) = t(\varphi(v_{i}))
\end{displaymath}
Examples of complete structural properties include $t(v_{i})=$ degree of node $v_{i}$,
$t(v_{i})=$ number of nodes at distance $k$ from $v_{i}$ ($k$-hop neighbors),
$t(v_{i})=$ the centrality of $v_{i}$,
etc.~\citep{batagelj1992direct}.
\end{definition}

Then the following theorem can be directly derived from~\citep{batagelj1992direct}:
\begin{theorem}
\label{theo:struc_iden}
Given a structural description of node $v_{i}$ defined by a set of complete structural node
properties as:
\begin{displaymath}
\bm{T}_{i} = [t_{1}(v_{i}), t_{2}(v_{i}),\cdots,t_{n}(v_{i})]
\end{displaymath}
where $\bm{T}_{i}$ is an $n$ dimensional vector, and
$n > 0$ is the number of chosen properties.
Let $d(\cdot,\cdot)$ denote standard dissimilarities between vectors, and
$v_{i}\equiv v_{j}$  indicate node $v_{i}$ and $v_{j}$ have equal structural identity,
then for $\forall v_{i},v_{j} \in V$:
\begin{displaymath}
v_{i}\equiv v_{j} \iff d(\bm{T}_{i}, \bm{T}_{j}) = 0
\end{displaymath}
\end{theorem}

From Theorem~\ref{theo:struc_iden}, we can conclude
that the structural identities between
nodes can be measured through properly designed structural vectors. Next we
are going to show that $\bm{S}^{\textrm{RPR}}_{i}$ actually represents a
complete
structural property.

Following the examples in Definition~\ref{def:node_proper}, the centrality
of $v_{i}$ can be regarded as a complete node property of $v_{i}$.
There are multiple ways to measure the centrality of a node,
while the original 
PageRank~\citep{brin2012reprint} is exactly a variant of eigenvector centrality.
At each iteration, the PageRank value $\bm{\pi}_{i}$ of node $v_{i}$
is updated as in~\citep{langville2011google}:
\begin{equation*} \small
\bm{\pi}_{i}^{T} = \bm{\pi}_{i}^{T} \bm{M},
\end{equation*}
and $\bm{M}$ is the Google matrix defined as:
\begin{equation}\small
\bm{M} = \beta \bm{P} + (1 - \beta)\bm{g}_{i}\bm{1}^{\textrm{T}}
\end{equation}
where $\bm{g}_{i}=(\frac{1}{|V|}, \cdots, \frac{1}{|V|})_{|V|}$
and $\bm{1} = (1, \cdots, 1)_{|V|}$.
According to Equation~(\ref{def:RPR}), as a variant of PageRank, 
the only difference between $\bm{S}_{i}^{\textrm{RPR}}$ and
the original PageRank
is the choice of $\bm{g}_{i}$. Therefore, the target matrix
of Rooted PageRank can be written as:
\begin{equation}\small
\bm{M}^{\textrm{RPR}}_{i} = \beta_{\textrm{RPR}}\bm{P} + (1 - \beta_{\textrm{RPR}}) \bm{I}
\end{equation}
where $\bm{I}$ is the identity matrix. Thus $\bm{S}_{i}^{\textrm{RPR}}$
is the leading left hand eigenvector of $\bm{M}^{\textrm{RPR}}_{i}$, i.e., $\bm{S}_{i}^{\textrm{RPR}}$
satisfies: $(\bm{S}_{i}^{\textrm{RPR}})^{T} = (\bm{S}_{i}^{\textrm{RPR}})^{T}\bm{M}^{\textrm{RPR}}_{i}$.
As a consequence, $\bm{S}_{i}^{\textrm{RPR}}$ is also a variant of
eigenvector centrality, thus
can be further regarded as a complete structural property
to represent the structural identity of $v_{i}$, i.e., $\bm{T}_{i} = \bm{S}^{\textrm{RPR}}_{i}$.

\subsection{Structural Feature Generation}

In this section we first state our motivation of choosing $\bm{S}^{\textrm{RPR}}_{i}$ as the structural
identity instead of others, based on which the structural feature generation method is introduced.

To construct an inductive method, we expect the length of the structural description of a node
is independent of the total number of nodes $|V|$
and fixed at $k$.{\tiny } 
In this paper, we use the top $k$ values of $\bm{S}^{\textrm{RPR}}_{i}$
as the structural description of $v_{i}$.
Compared with other high-order proximities or node properties (e.g., the original PageRank),
RPR captures the global structural information of the network, while being tailored to encode the local structural information of root nodes according to the definition, and thus can better represent the importance of neighbors at various distances to the present node~\citep{haveliwala2002topic}.
Moreover,
it has been theoretically and empirically proven~\citep{litvak2007degree}
that in a network with power-law degrees,
the RPR
values also follow a power-law distribution.
Therefore,
the largest $k$ RPR values are able to
represent the structural
information of a node with a suitably chosen $k$.

\begin{algorithm}[tb] 
\caption{Rooted random walk sampling}
\label{alg:rootedsam}
\begin{algorithmic}[1]
\REQUIRE the graph $G$,
the present node $v_{i}$, the continuation probability $\beta_{\textrm{RPR}}\in (0,1)$,
hyper-parameters $k$, $m$, $l$
\ENSURE the structural feature vector $\bm{T}_{i}$ of $v_{i}$
\STATE Initialize a counter $C_{i} \in \mathbb{R}^{|V|}$
\REPEAT
\STATE $P_{s}$ = RootedRandomWalk($G$, $v_{i}$, $l$, $\beta_{\textrm{RPR}}$)
\FOR {$v_{j}$ in $P_{s}$}
\STATE $C_{i}[j] \gets C_{i}[j] + 1$
\ENDFOR
\UNTIL{$m$ times}
\STATE $C_{i}\gets$ Sort $C_{i}$ in a descending order
\STATE $\bm{T}_{i}\gets$ The first $k$ elements of $C_{i}$
\RETURN $\bm{T}_{i}\gets \bm{T}_{i} / \mathrm{sum}(\bm{T}_{i})$
\end{algorithmic}
\end{algorithm}

When calculating $\bm{S}_{i}^{\textrm{RPR}}$, considering the \emph{inductive} prerequisite,
the transition matrix $\bm{P}$ which encodes the structure 
of the network may be unavailable. 
Alternatively,
we approximate $\bm{S}_{i}^{\textrm{RPR}}$ from the local structure around $v_{i}$
through a Monte Carlo approximation. 
The complete procedure
is summarized in Algorithm~\ref{alg:rootedsam}, where $m$ and $l$ indicate
the number of repeats and the length per random walk respectively, and $k$ controls the length of
the structural feature vector.
The largest $k$ values of $\bm{S}^{\textrm{RPR}}_{i}$ are taken
as the structural description of $v_{i}$, denoted as $\bm{T}_{i}$ in the
rest of this paper.

\subsection{Embedding Generation}
\label{sec:emb_gen}

To construct an inductive network embedding framework, we generate embeddings
instead of directly
optimizing the embedding matrix, through which the structural information and the
content information of networks can be jointly incorporated.

As stated above, 
the $k$ values in $\bm{T}_{i}$ indicate $k$ largest structural proximities between $v_{i}$ and nodes
co-occurring with $v_{i}$ in rooted random walks. We denote the content matrix of
the corresponding $k$ nodes as $\bm{F}_{i}^{k} \in \mathbb{R}^{k \times f}$, which is
constructed row by row according to the same order of RPR values in $\bm{T}_{i}$.\
Given the structural features $\bm{T}_{i}$ and node content $\bm{F}_{i}^{k}$,
we propose an embedding generation method for $v_{i}$.

Specifically, we first employ a multilayer perceptron~(MLP)
to map nodes from the content space to the embedding space, then
compute a linear combination of the $k$ vectors with respect to
the corresponding weights in $\bm{T}_{i}$. Formally, denote the dimensionality of
embeddings as $d$, the weight matrix of the MLP is
$\bm{W_{\textrm{M}}} \in \mathbb{R}^{f \times d}$, then the embedding generation process
can be written as:
\begin{equation} \small
\label{equ:embedding}
\bm{e}_{i} = \sigma(\sum_{j=1}^{k} T_{i,j} \bm{F}_{i,j}^{k} \bm{W_{\textrm{M}}}),
\end{equation}
where $\bm{F}_{i,j}^{k} \in \mathbb{R}^{f}$ is the $j$-th row of $\bm{F}_{i}^{k}$,
$\sigma$ is the non-linear activation function, and
$\bm{e}_{i} \in \mathbb{R}^{d}$ is the embedding vector of node $v_{i}$.

\subsection{Biased SGNS}
\label{sec:biased}

The Skip-Gram Negative Sampling (SGNS) model is widely used in representation learning, which is based on the principle of learning the embedding vector of a word/node by predicting its neighbors.
More formally, given an embedding vector $\bm{e}_{i}$, SGNS is minimizing the objective function as:
\begin{equation} \small
\label{equ:SGNS} 
J(\bm{e}_{i} \mid \bm{e}_{p}) = -\textrm{log}(\sigma(\bm{e}^{T}_{i} \bm{e}_{p})) - K \cdot \mathbb{E}_{v_{n}
\sim P_{n}(v_{p})}\textrm{log}(\sigma(-\bm{e}^{T}_{i} \bm{e}_{n})),
\end{equation}
where $\sigma(\cdot)$ is the sigmoid function and $K$ controls the negative sampling number.
$v_{p}$ and $v_{n}$ are positive and negative nodes of $v_{i}$ respectively, while $\bm{e}_{p}$
and $\bm{e}_{n}$ are the corresponding embedding vectors. 
Technically,
negative nodes are sampled from a
distribution $P_{n}$, and for most network
embedding methods,
positive nodes are defined as nodes that co-occur
with $v_{i}$ in a fixed-size window in random walk sequences.

In SPINE, 
to encourage the similarity of embeddings to jointly
encode the similarity in terms of structural identities and local proximities simultaneously,
we 
design a novel biased positive sampling strategy based on the structural features
generated from Algorithm~\ref{alg:rootedsam}. The complete procedure is illustrated in
Algorithm~\ref{alg:biased}.
Specifically, we define a structural rate $\alpha \in (0,1)$ to control the ratio of
structural sampling and local proximity sampling. With probability $\alpha$, 
a positive sample of $v_{i}$ is sampled
according to the similarities between their structural
features (starting from line $2$).
Otherwise, the positive node is sampled from
nodes that co-occur near $v_{i}$ on trivial random walks (starting from line $9$),
which is preprocessed and stored in $L_{i}$.
The similarity metric in line $4$ can be chosen from Euclidean distance, cosine similarity, Dynamic Time Warping (DTW)~\citep{salvador2007toward}, etc.
In our experiments, we use DTW which is designed to compare ordered sequences as the similarity metric. 

\begin{algorithm}[tb] 
\caption{Biased positive sampling}
\label{alg:biased}
\begin{algorithmic}[1]
\REQUIRE the structural feature matrix $\bm{T}$, the present node $v_{i}$,
a node list $L_{i}$,
the structural rate $\alpha$
\ENSURE $v_{p}$, which is a positive sample of $v_{i}$
\STATE Initialize an empty list $P_{s}=[]$
\IF {$\textrm{random}(0,1) < \alpha$}
\FOR {$j=1$ to $|V|$, $j \neq i$}
\STATE $t\gets$ Compute the similarity between $\bm{T}_{i}$ and $\bm{T}_{j}$
\STATE $P_{s}\gets$ Append $t$ to $P_{s}$
\ENDFOR
\STATE $P_{s} \gets$ Normalize $P_{s}$ to $[0,1]$
\STATE $v_{p} \gets$ Sample a node according to $P_{s}$
\ELSE
\STATE $v_{p} \gets$ Randomly choose a node from $L_{i}$
\ENDIF
\RETURN $v_{p}$
\end{algorithmic}
\end{algorithm}

The structural sampling paradigm alleviates the limitation of distance.
In practice, it is redundant to compute the structural similarity between $v_{i}$
and all the other nodes, since nodes with completely different local structures are
nearly impossible to be sampled as positive pairs through structural sampling.
Intuitively, nodes with
similar degrees are likely to have similar local structures. Based on this intuition,
we reduce the redundancy by only
considering nodes which have similar degrees with the present node $v_{i}$.
Specifically, given an ordered list of node degrees, we choose the candidates for structural sampling
by taking $O(\log|V|)$ nodes in each side from the location of $v_{i}$.
As a consequence, the time complexity of structural sampling for each node is
reduced from $O(|V|)$ to $O(\log|V|)$.

\subsection{Learning and Optimization}
\label{sec:opt}

We introduce the biased SGNS based objective function of our framework in this section.
We propose two types of embeddings for each node $v_{i} \in V$, i.e., a content
generated embedding $\bm{e}_{i}$ as defined in Equation~(\ref{equ:embedding}), and
a structure based embedding $\bm{s}_{i}$ which is the $i$-th row of an auxiliary embedding
matrix $\bm{W}_{\textrm{S}} \in \mathbb{R}^{|V| \times d}$.
In real-world network-based datasets,
the content of nodes is likely to be extremely sparse and weaken the structural information
incorporated during the generation process of $\bm{e}_{i}$. Therefore we employ
a direct interaction between $\bm{e}_{i}$ and $\bm{s}_{i}$ to strengthen the
structural information contained in the learned embeddings.

Formally, given the present node $v_{i}$ and its positive sample $v_{j}$,
which is sampled according to Algorithm~\ref{alg:biased},
the pairwise objective function of SPINE can be written as:
\begin{equation} \small
\begin{aligned} 
\label{equ:obj0}
F(v_{i}, v_{j}) &= \lambda_{1} \cdot J(\bm{e}_{i} \mid \bm{e}_{j}) + \lambda_{2}\cdot J(\bm{s}_{i}\mid \bm{s}_{j}) \\
&+ (1-\lambda_{1} - \lambda_{2})\cdot [J(\bm{e}_{i}\mid \bm{s}_{j}) + J(\bm{s}_{i}\mid \bm{e}_{j})]
\end{aligned}
\end{equation}
where $\lambda_{1}$ and $\lambda_{2}$ control weights of different parts,
and $J(\cdot | \cdot)$ denotes the pairwise SGNS between two embeddings defined
in Equation~(\ref{equ:SGNS}).
Intuitively, the generator and the auxiliary embedding matrix
should be well trained through their single loss as well as obtaining each other's
information through the interaction loss, where $\lambda_{1}$ and $\lambda_{2}$
determine which one is the primary part.

The structure based embeddings $\bm{W}_{\textrm{S}}$ and the parameters $\bm{W}_{\textrm{M}}$ of the MLP
constitute all the parameters to be learned in our framework. 
The final objective of our framework is:
\begin{equation} \small
\label{equ:obj}
\min_{\bm{W}_{\textrm{S}}, \bm{W}_{\textrm{M}}} 
\sum_{ i \neq j}^{|V|}.
F(v_{i}, v_{j})
\end{equation}

By optimizing the above objective function,
the embedding generator 
is supposed to contain the content information and the structural
information as well as local proximities and structural identities
simultaneously. Therefore, during inference, we drop the
auxiliary embedding matrix $\bm{W}_{\textrm{S}}$ and only keep the trained embedding
generator.
In the sequel, embeddings of unseen nodes can be generated by first
constructing structural features via
Algorithm~\ref{alg:rootedsam} and then following the paradigm described in Section~\ref{sec:emb_gen}.

\section{Experiments}
\label{sec:exp}

\subsection{Experimental Setup}

\begin{table}[tb]
\centering \small
\begin{tabular}{c||ccc}
\hline
Method & Citeseer & Cora & Pubmed \\
\hline
\hline
node2vec & $47.2$ & $69.8$ & $70.3$ \\
struc2vec & $41.1$ & $64.2$ & $60.7$ \\
\emph{n+s} & $42.7$ & $68.3$ & $67.1$ \\
\emph{n+s+f} & $57.0$ & $73.7$ & $67.3$ \\
SDNE & $45.2$ & $68.7$ & $69.1$ \\
HOPE & $46.1$ & $67.2$ & $69.4$ \\
Graphsage & $52.6$ & $79.8^{\dagger}$ & $75.2$ \\
GAT & $72.5^{\dagger} \pm 0.7$ & $83.0^{\dagger} \pm 0.7$ & $79.0^{\dagger} \pm 0.3$ \\
\hline
SPINE & $72.6 \pm 0.4$ & $\bm{83.7} \pm 0.4$ & $78.5 \pm 0.3$ \\
SPINE-p & $\bm{73.8} \pm 0.2$ & $82.2 \pm 0.7$ & $\bm{82.2} \pm 0.3$ \\
\hline
\end{tabular}
\vskip -0.1in
\caption{Accuracy of transductive node classification (in percentage). ``$^{\dagger}$'' indicates that the results are directly copied from their papers and other results are provided by ourselves.
}
\label{tab:trans_node}
\vspace{-15pt}
\end{table}

We test the proposed model on four benchmark datasets to measure its performance
on real-world tasks, and one small scale social network to validate
the structural identity preserved in the learned embeddings.
For the node classification task, we test our method on Citation Networks~\citep{yang2016revisiting},
where nodes and edges represent papers and citations respectively.
To test the performance of SPINE while generalizing across networks,
we further include PPI~\citep{stark2006biogrid}, which consists of multiple networks
corresponding to different human tissues. 
To measure the structural identity preserved in embeddings, we test SPINE on a subset
of Facebook dataset~\citep{snapnets}, denoted as FB-686, in which nodes and links represent
users and their connections, and each user is described by a binary vector.

As for the baselines, we consider unsupervised network embedding methods including
node2vec~\citep{grover2016node2vec}, struc2vec~\citep{ribeiro2017struc2vec} and
their variants. 
Considering that node2vec and struc2vec are designed to preserve the local proximity
and the structural identity respectively, we concatenate the corresponding
learned embeddings
to form a new baseline, denoted as \emph{n+s}, to illustrate the superiority of SPINE
over the linear combination of the two proximities. In addition, \emph{n+s+f} denotes the content-incorporated variant of \emph{n+s}.
We also compare with SDNE~\citep{wang2016structural} and HOPE with RPR matrix~\citep{ou2016asymmetric} to test the performance of our
inductive Rooted PageRank approximation.
On the other hand, in addition to transductive methods, we also consider the unsupervised variant of Graphsage~\citep{hamilton2017inductive}, an inductive network embedding
method which jointly leverages structural and content information.
We also report the performance of the state-of-the-art supervised inductive node classification
method GAT~\citep{petar2017graph}.
Random and raw feature results are also included as baselines in this setting.

For our method, we use SPINE and SPINE-p to indicate the variants with $\bm{W}_{\textrm{S}}$ randomly initialized and pretrained with node2vec respectively.
To make predictions based on the embeddings learned by unsupervised models,
we use one-vs-rest logistic regression as the downstream classifier.
For all the methods, the dimensionality of embeddings is set to $200$\footnote{Code is avaliable at \url{https://github.com/lemmonation/spine}}.

\subsection{Node Classification}
We first evaluate the performance of SPINE on node classification,
a common network mining task. Specifically, we conduct the experiments in both
transductive and inductive settings. 
For the transductive setting,
we use the same scheme of training/test partition
provided by~\citet{yang2016revisiting}.
As for the inductive setting, on citation networks,
we randomly remove $20\%$, $40\%$, $60\%$ and $80\%$ nodes
and the corresponding edges, 
these nodes are then treated as test nodes 
with the remaining network as the training data.
Meanwhile on the PPI network, we follow the same dataset splitting strategy as in~\citep{hamilton2017inductive},
i.e.,  20 networks for training, 2 for validation
and 2 for testing, where the validation and testing networks remain unseen during training.
For both settings we repeat the process 10 times and report the mean score.



\begin{figure}[tb]
\centerline{\includegraphics[width=0.8\columnwidth]{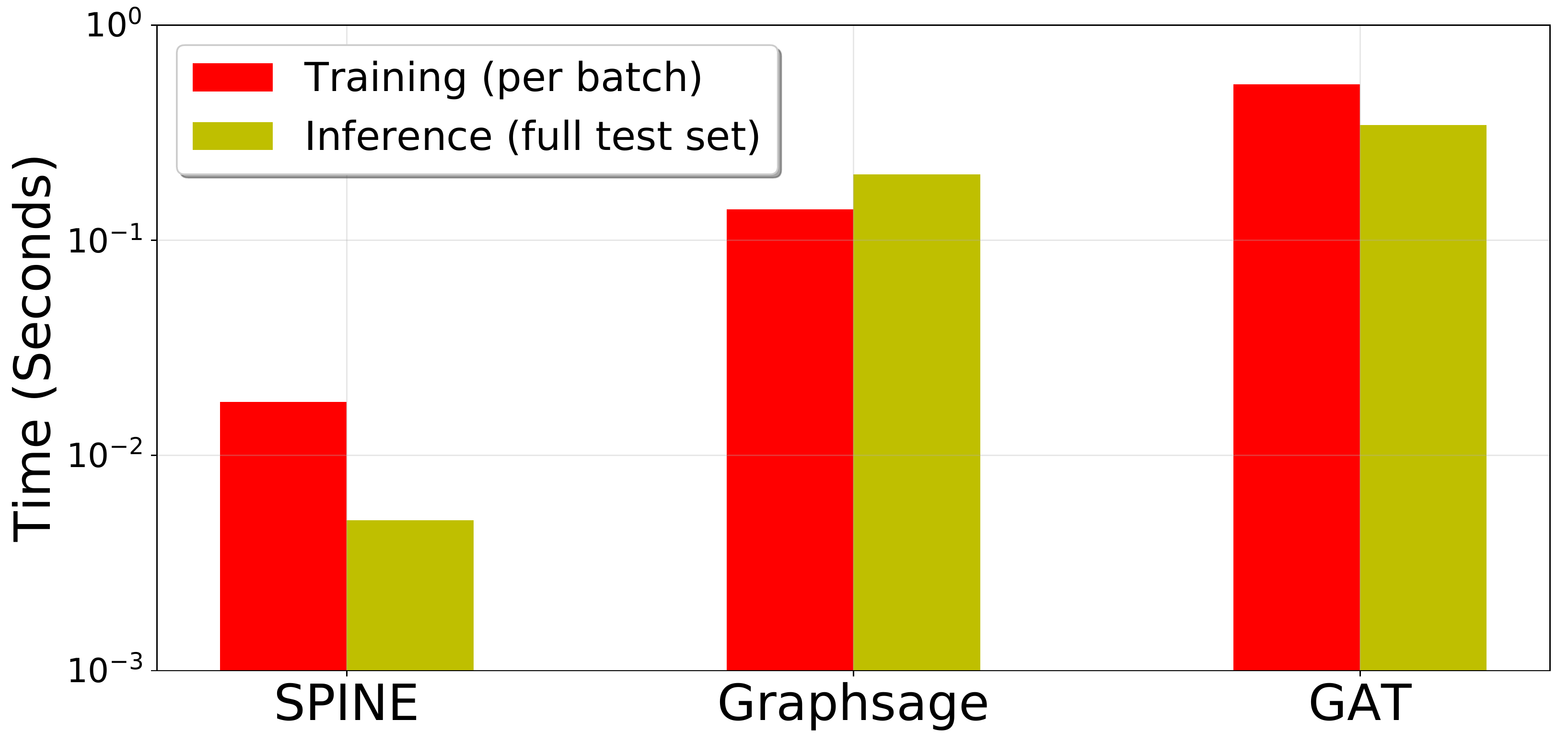}}
\vskip -0.1in
\caption{
Comparison of running time on Cora, with training batches of size 512 and inference on the full test set (1000 nodes).
}
\label{fig:time_bar}
\vspace{-10pt}
\end{figure}

Results in the \textbf{transductive} setting are reported in Table~\ref{tab:trans_node}. We can observe that SPINE outperforms
all unsupervised embedding methods, and performs comparably with the state-of-the-art supervised
framework GAT. In addition, \emph{n+s} performs worse than node2vec, which implies that
a simple linear combination of local proximity preserved and structural identity preserved embeddings
is incapable of generating a meaningful representation that effectively integrates the two components. 
The superiority of SPINE over SDNE and HOPE indicates the efficacy of the inductive RPR approximation algorithm as well as
the joint consideration
of local proximity and structural identity.
SPINE also outperforms the content-augmented variant \emph{n+s+f},
which shows that
the content aggregation method we propose can better consolidate the content and structure information. Furthermore, the comparison between the two variants of SPINE indicates that while the basic model of SPINE already achieves a competitive performance, we can further enhance the model with initializations of $\bm{W}_{S}$ that are well pretrained by focusing only on local proximity, which also justifies the effectiveness of the paradigm of interactive integration proposed in Section~\ref{sec:opt}.


As for the comparison on training and test runtime, results are
shown in Figure~\ref{fig:time_bar}.
Obviously,
SPINE is more efficient in time complexity, especially in the inference stage.
In addition, SPINE is also more compact in space complexity, as
the parameter scale of SPINE during inference is $O(fd)$, compared to $O(fd+d^{2})$ and $O((f+KC)d)$ for
Graphsage and GAT with two layers respectively, where $K$ is the number of attention heads and $C$ is
the number of classes. 


\begin{table}[tb]
\centering \small
\begin{tabular}{c||c|c|c|c|c}
\hline
&Methods&$20\%$ & $40\%$ & $60\%$ & $80\%$\\
\hline
\hline
\multirow{4}{*}{Citeseer} &
Random &  $19.5$ & $20.4$ & $16.7$ & $17.7$\\
&RawFeats &  $63.9$ & $62.2$ & $60.3$ & $57.7$\\
&Graphsage &  $58.5$ & $53.9$ & $47.8$ & $41.4$\\
\cline{2-6}
&SPINE &  $\bm{75.4}$ & $\bm{72.1}$ & $\bm{71.5}$ & $\bm{68.7}$\\
\hline
\multirow{4}{*}{Cora} &
Random &  $18.8$ & $22.0$ & $19.1$ & $20.1$\\
&RawFeats &  $66.6$ & $64.7$ & $64.6$ & $59.6$\\
&Graphsage &  $73.1$ & $66.4$ & $58.8$ & $48.6$\\
\cline{2-6}
&SPINE &  $\bm{86.7}$ & $\bm{84.1}$ & $\bm{82.1}$ & $\bm{77.9}$\\
\hline
\multirow{4}{*}{Pubmed} &
Random &  $38.5$ & $39.8$ & $39.3$ & $38.9$\\
&RawFeats &  $75.7$ & $75.4$ & $74.6$ & $72.9$\\
&Graphsage &  $79.9$ & $79.4$ & $78.2$ & $76.4$\\
\cline{2-6}
&SPINE &  $\bm{85.7}$ & $\bm{83.7}$ & $\bm{83.0}$ & $\bm{78.8}$\\
\hline
\end{tabular}
\vskip -0.1in
\caption{Accuracy of inductive node classification w.r.t node removal rate (in percentage).}
\label{tab:ind_node}
\vspace{-10pt}
\end{table}

\begin{figure}[tb]
\centerline{\includegraphics[width=0.8\columnwidth]{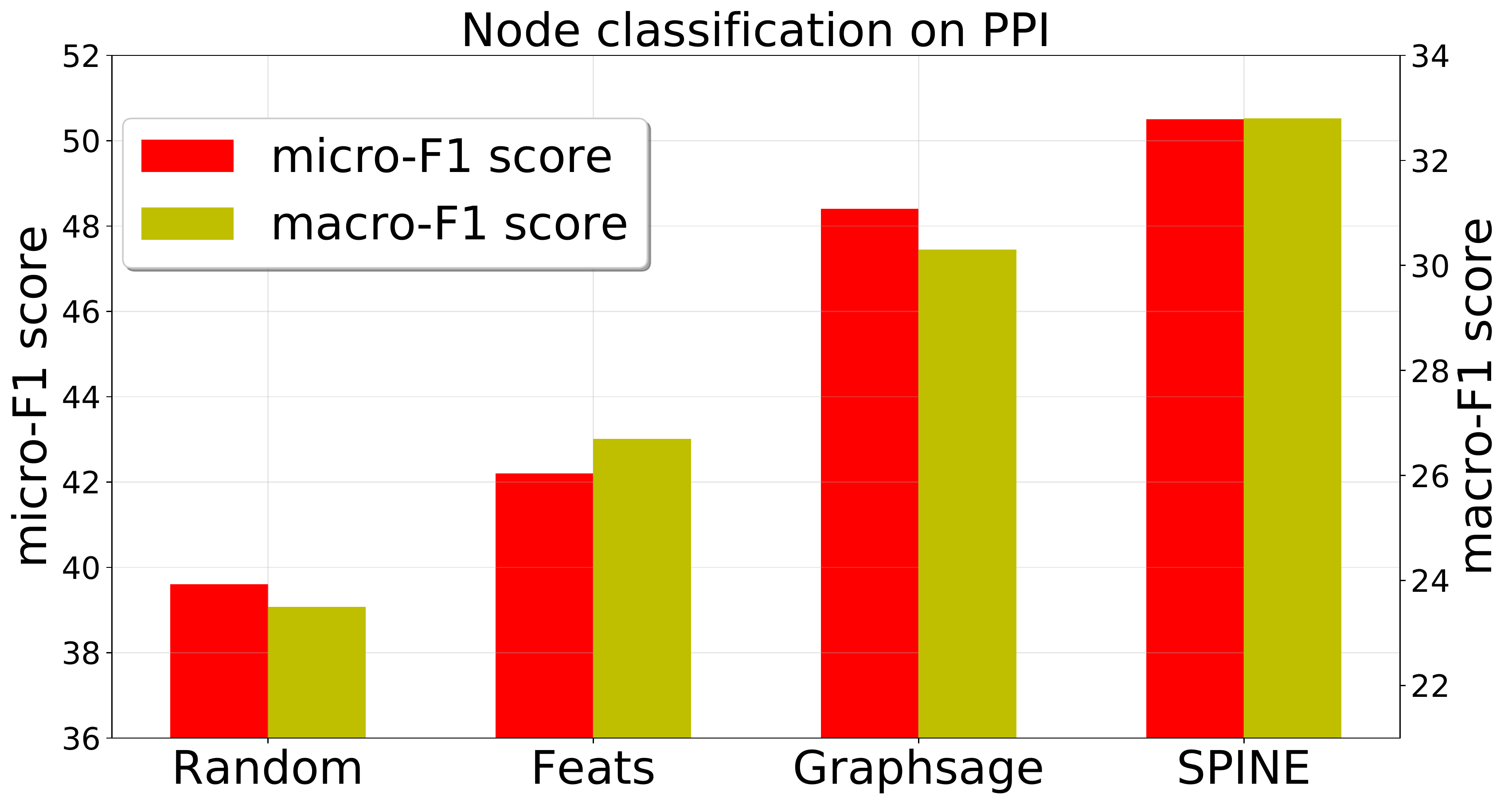}}
\vskip -0.1in
\caption{
Node classification results on PPI.
The left vertical axis indicates the micro-F1 score while the right indicates
the macro-F1 score. Both are in percentage.
}
\label{fig:ppi_bar}
\vspace{-10pt}
\end{figure}

Results in the \textbf{inductive} setting are reported in Table~\ref{tab:ind_node} and
Figure~\ref{fig:ppi_bar}.
Note that in this setting we use all the remaining nodes as training data during classification, thus the results are generally larger than that under the transductive setting.
One can observe that SPINE outperforms all the baselines, indicating the generalization
capability of the embedding generator
learned by optimizing our carefully designed objective function.
In addition, with the increasing node removal rate which leads to 
greater loss of local proximities, Graphsage can perform worse than raw features, indicating the limitation of the methods that only preserve local proximities.
In contrast, SPINE alleviates the sparsity of
local information by incorporating structural identities.

\subsection{Structural Identity}

We proceed to investigate the structural identity 
on the FB-686 dataset here.
We consider the transductive setting here, and the results under inductive setting can be found in the supplementary material.
Specifically, for the original network, we construct a mirror network and
relabel the nodes, and consider the union of two networks as the input.
As a consequence,
node pairs between original nodes and their mirror nodes are obviously structurally equivalent,
thus should be projected close in the embeddings space.
We then evaluate the Euclidean distance distribution between embeddings of the mirror node pairs
and all the node pairs connected by edges,
denoted as $P_{m}$ and $P_{a}$ respectively. Intuitively, if embeddings successfully preserve
structural identities of nodes, $\mathbb{E} [P_{m}]$ should be much smaller than $\mathbb{E}[P_{a}]$.

\begin{figure}[tb]
\centerline{\includegraphics[width=0.8\columnwidth]{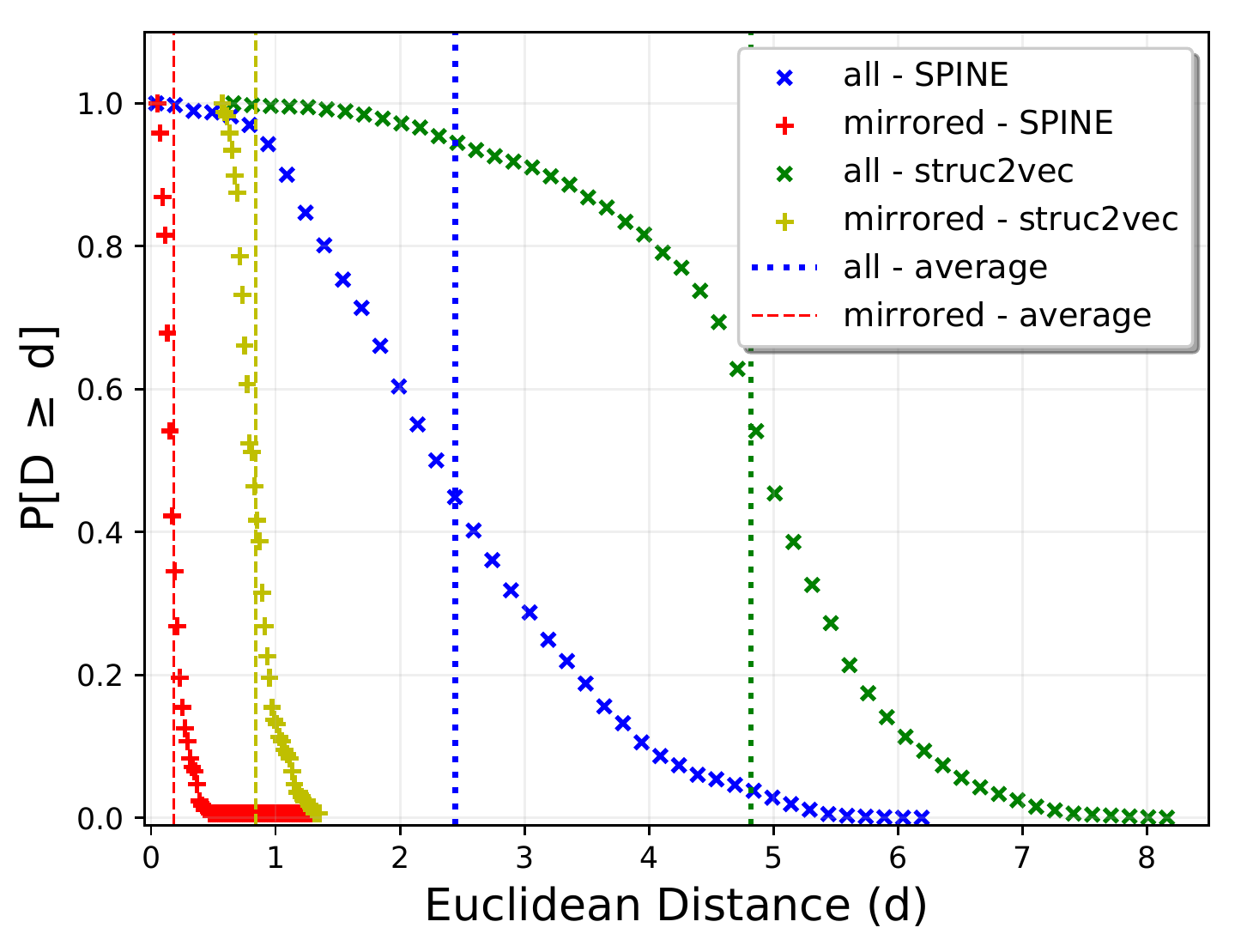}}
\vskip -0.1in
\caption{
Euclidean distance distribution between mirrored node pairs and connected node pairs on the FB-686 dataset.
}
\label{fig:compare_dist}
\vspace{-10pt}
\end{figure}

Results of SPINE and struc2vec with respect to the two distributions are shown in Figure~\ref{fig:compare_dist}. 
Obviously, compared to struc2vec, 
embeddings learned by SPINE yield smaller distances 
between both mirrored node pairs and connected node pairs, indicating the structural identity and local proximity are
jointly preserved better. 
In addition, the ratio between $\mathbb{E}[P_{a}]$ and $\mathbb{E} [P_{m}]$ is $13.40$ and $5.72$ for
SPINE and struc2vec respectively, which means SPINE distinguishes the two proximities more clearly.

\begin{figure}[tb]
\centerline{\includegraphics[width=0.8\columnwidth]{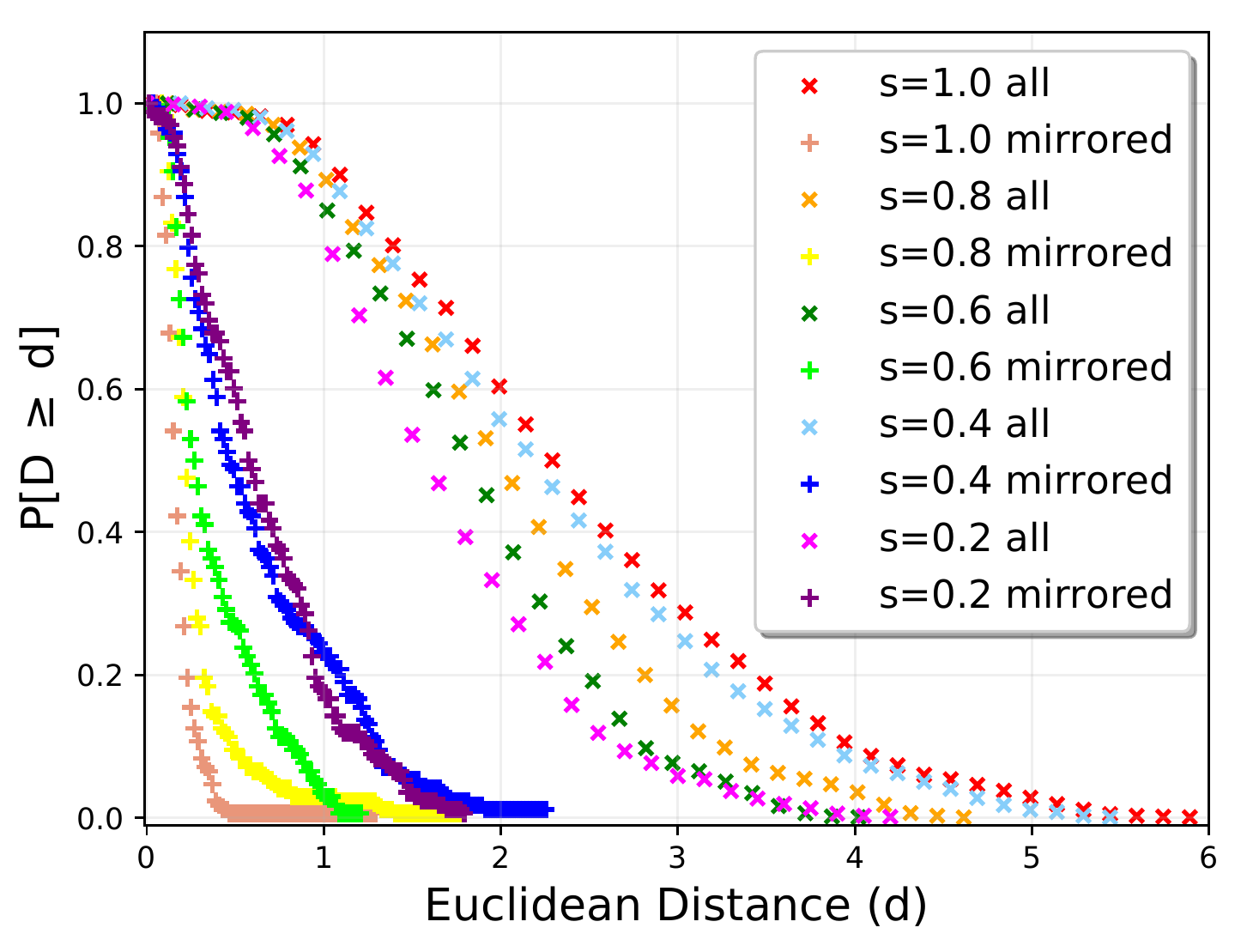}}
\vskip -0.1in
\caption{
Euclidean distance distribution between mirrored node pairs and connected node pairs on the FB-686 datasets with varying degrees of edge removal and changing content.
}
\label{fig:res_dist}
\vspace{-10pt}
\end{figure}

Further to test the robustness of SPINE to edge removal and changing content, we randomly sample two
new networks from the original FB-686 network. Specifically, we preserve each edge in
the original network with probability $s$, and randomly exchange a $1$'s location with another $0$'s location
in each node's content vector. Consequently, from the view of structure,
the probability for an original edge contained
both in the two generated networks is $s^{2}$, and smaller $s$ indicates less structure correlation
between the two generated networks. From the view of content, mirrored nodes are nearly impossible
to have identical content due to the sparsity of content vectors.
As can be observed in Figure~\ref{fig:res_dist}, the ratio between
$\mathbb{E}[P_{a}]$ and $\mathbb{E} [P_{m}]$ is not significantly affected by the degree of structure perturbation $s$, which indicates that SPINE can robustly distinguish and preserve
structural identity as well as local proximity even with severe perturbations.

\section{Related Work}
Most network embedding methods consider to preserve local proximity between nodes with frameworks based on
random walk~\citep{perozzi2014deepwalk,grover2016node2vec}, skip-gram~\citep{tang2015line,cao2015grarep}, matrix factorization~\citep{yang2015network,guo2017enhancing} and deep learning~\citep{wang2016structural,gao2018deep} respectively.
However, it is worth noting that few of the existing works consider the
structural identity between nodes, and fail to handle proximities between
nodes at distances. 
Struc2vec~\citep{ribeiro2017struc2vec} preserves the structural identity by constructing a multi-layer complete graph and execute random walk on it. HOPE~\citep{ou2016asymmetric} captures structural identity through factorizing a global Rooted PageRank matrix. However, while preserving the structural identity, they ignore the basic local proximities of nodes, which limits its applicability on real-world network mining tasks.
Similar problems also occur in two recent methods~\citep{tu2018deep,zhang2018arbitrary}. SDNE~\citep{wang2016structural}, a deep learning based method, is only able to take the first- and second-order proximities into account.
Furthermore, most of the methods mentioned above are transductive. Inductive methods~\citep{hamilton2017inductive,petar2017graph} tackles this challenge by recursively training a set of aggregators for each node to integrate its neighbors' content as the embedding of the current node in every iteration. As nodes at the $k$-th iteration contain the structural information from their neighbors within $k$ hops, they cannot deal with nodes at arbitrary distances unless with sufficient iterations, which is costly for real-world tasks.

\section{Conclusion}
\label{sec:conclu}
In this paper, we propose SPINE, a network embedding approach
which is able to jointly preserve structural identities and local proximities
of nodes while being generalized
to unseen nodes or networks. 
We assign a structural feature vector to each node based on Rooted PageRank,
and we learn an embedding generator leveraging the structural features of each node
to incorporate the structural and content information
of nearby nodes. 
In addition, 
we propose
a biased SGNS algorithm with a novel positive
sampling procedure, based on which a carefully designed objective function
is proposed to enhance the structural information contained in the embedding generator.
Extensive experiments demonstrate the superiority
of SPINE over the state-of-art baselines
on both transductive and inductive
tasks.
In future work, we are interested in introducing structural identity to other network-based tasks such as social recommendation.

\section*{Acknowledgements}
This research was supported by the National Natural Science Foundation of China (No. 61673364, No. 91746301) and the Fundamental Research Funds for the Central Universities (WK2150110008).

\bibliographystyle{named}
\bibliography{ijcai19}

\appendix

\section{Rooted Random Walk}

Please refer to Algorithm~\ref{alg:rooted} for details of the Monte Carlo approximation of rooted random walk.

\begin{algorithm}[tb] \small
\caption{RootedRandomWalk}
\label{alg:rooted}
\begin{algorithmic}[1]
\REQUIRE the graph $G$, the present node $v_{i}$,
the continuation probability $\beta_{\textrm{RPR}}\in (0,1)$, the walk length $l$
\ENSURE a rooted random walk sequence $P_{s}$
\STATE Initialize a list $P_{s} = [v_{i}]$
\FOR {$j=1$ to $l - 1$}
\STATE $v_{cur} \gets$ the last element of $P_{s}$
\IF {$\textrm{random}(0,1) < \beta_{\textrm{RPR}}$}
\STATE Randomly select a node $v_{j}$ from the neighbors of $v_{cur}$
\STATE Append $v_{j}$ to the end of $P_{s}$
\ELSE
\STATE Append $v_{i}$ to the end of $P_{s}$
\ENDIF
\ENDFOR
\RETURN $P_{s}$
\end{algorithmic}
\end{algorithm}

\section{Dataset Details}

\begin{table}[tb]
\centering
\begin{tabular}{ccccc}
\hline
Dataset&\# Classes&\# Nodes&\# Edges&\# Features\\
\hline
\hline
Citeseer & $6$ & $3327$ & $4732$ & $3703$ \\
Cora & $7$ & $2708$ & $5429$ & $1433$ \\
Pubmed & $3$ & $19717$ & $44338$ & $500$ \\
PPI & $121$ & $56944$ & $818716$ & $50$ \\
FB-686 & - & $168$ & $3312$ & $63$ \\
\hline
\end{tabular}
\caption{Dataset statistics}
\label{tab:data}
\vskip -0.1in
\end{table}

We test the proposed model on four benchmark datasets. As most existing methods are transductive,
we adopt three static datasets and one across network dataset to measure the transductive and inductive performance
of SPINE respectively. 
The statistics of datasets
are summarized in Table~\ref{tab:data}. 
Among them, the static datasets are Citation Networks:

\paragraph{Citeseer}
contains 3312 publications of 6 different classes and 4732 edges between
them. Nodes represent papers
and edges represent citations. Each paper is described by a one-hot vector of 3703
unique words.

\paragraph{Cora} includes 2708 publications
of machine learning with 7 different classes. Similar
to Citeseer, Cora contains 5429 citation links between them. And each paper is described
by a one-hot vector of 1433 unique words.

\paragraph{Pubmed} consists of 19717 scientific publications
pertaining to diabetes classified into one of three classes. It contains 44338 citation links
and each document is described by a TFIDF weighted word vector from a dictionary
of 500 unique words.

To test the performance of SPINE while generalizing across networks, we further introduce the
PPI dataset:

\paragraph{PPI} contains various protein-protein interactions,
where each graph corresponds to a different human tissue. Nodes represent proteins
with $121$ different cellular functions from gene ontology as their labels.

\paragraph{FB-686} is a subset of Facebook dataset, which consists of
168 users with 3312 links between them. Each user is described by a 0-1 vector of
63 features.

\section{Hyperparameter Settings}
We keep the following settings for all tasks and datasets:
ratios $\lambda_{1}$ and  $\lambda_{2}$ 
are
set to $0.4$ and $0.2$ respectively, while the structural rate $\alpha$
and the restart rate $\beta_{\textrm{RPR}}$
of rooted random walk
are both set to $0.5$. The learning rate of the Adam optimizer is
set to $0.001$. For methods that leverage random walks,
we set the number of repeats for each node to $10$, the length of random walks to $40$
and the window size to $5$ for a fair comparison.

\section{Parameter Study}

In order to evaluate the influence of parameters on the performance of SPINE, we conduct experiments on
a subset of PPI datasets, denoted as subPPI,
which contains $3$ training networks and
$1$ test network. We first investigate the effect of the structural rate
$\alpha$,
then
test the robustness of SPINE with respect to varying $\lambda_{1}$ and
$\lambda_{2}$ values, both on node classification.

\subsection{Effect of $\alpha$}

\begin{figure}[tb]
\centerline{\includegraphics[width=0.8\columnwidth]{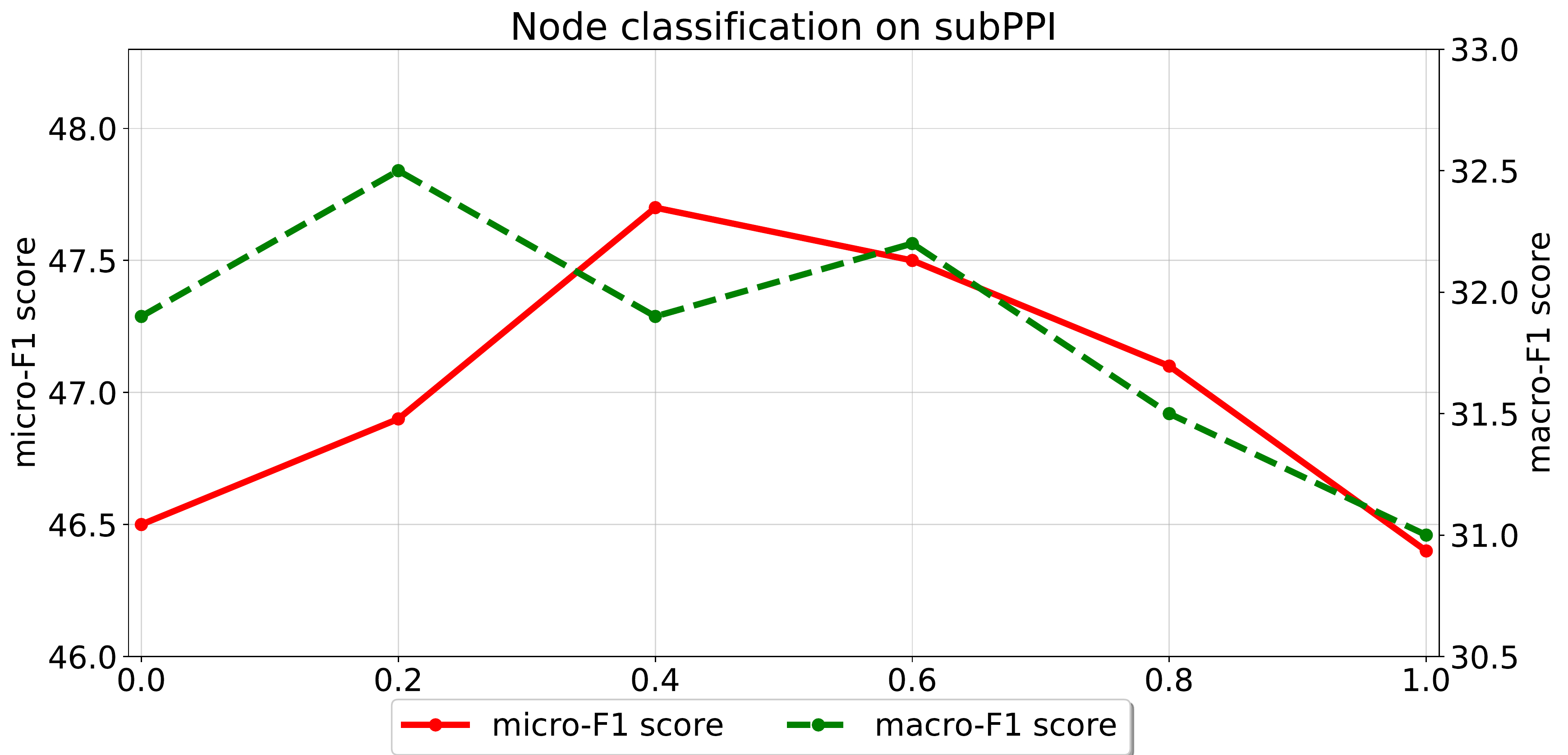}}
\vskip -0.1in
\caption{
Classification results with different $\alpha$ on subPPI.
The left and right vertical axes indicates micro-F1 and macro-F1 score respectively.
Both are in percentage.
}
\label{fig:subppi_kdd}
\vskip -0.1in
\end{figure}

We vary the value of the structural rate $\alpha$ from $0$ to $1$ with an interval of
$0.2$, and report the results in Figure~\ref{fig:subppi_kdd}. As expected,
with increasing $\alpha$ values, the classification
performance increases first and then decreases when $\alpha$ gets too large.
The results show that only considering the structural identity or local proximity
will impair the quality of learned embeddings in real-world tasks.
Therefore,
the structural rate $\alpha$ is crucial
for enhancing the quality of learned embeddings.

\subsection{Effects of $\lambda_{1}$ and $\lambda_{2}$}

\begin{table}[tb]
\centering
\caption{Results of node classification on subPPI with varying $\lambda_{1}$ and $\lambda_{2}$ values (in percentage)}
\label{tab:lambda_subppi}
\begin{tabular}{c|c||c|c}
\hline
$\lambda_{1}$&$\lambda_{2}$&micro-F1 & macro-F1\\
\hline
\hline
$0.2$ & $0.2$ & $47.1$ & $32.0$\\
$0.2$ & $0.4$ & $47.0$ & $31.4$\\
$0.2$ & $0.6$ & $46.6$ & $30.5$\\
$0.4$ & $0.2$ & $47.2$ & $\bm{32.1}$\\
$0.4$ & $0.4$ & $\bm{47.8}$ & $30.5$\\
$0.6$ & $0.2$ & $46.7$ & $31.2$\\
$1.0$ & $0.0$ & $45.4$ & $30.4$\\
\hline
\end{tabular}
\vskip -0.1in
\end{table}

To investigate the influence of ratios $\lambda_{1}$ and $\lambda_{2}$,
we vary the value of $\lambda_{1}$ from $0.2$ to $1.0$ and the value of $\lambda_{2}$ from $0.2$ to $0.6$,
both with an interval of
$0.2$. Results of node classification on subPPI are reported in Table~\ref{tab:lambda_subppi},
from which 
we can conclude that the interaction between
the generator and $\bm{W}_{\textrm{S}}$ successfully enhances the structure information
carried in the generator as expected,
as $\lambda_{1}=1.0$ (no interaction) achieves
the worst performance.

\section{Case Study}

\begin{figure*}[tb]
\centerline{\includegraphics[width=1.6\columnwidth]{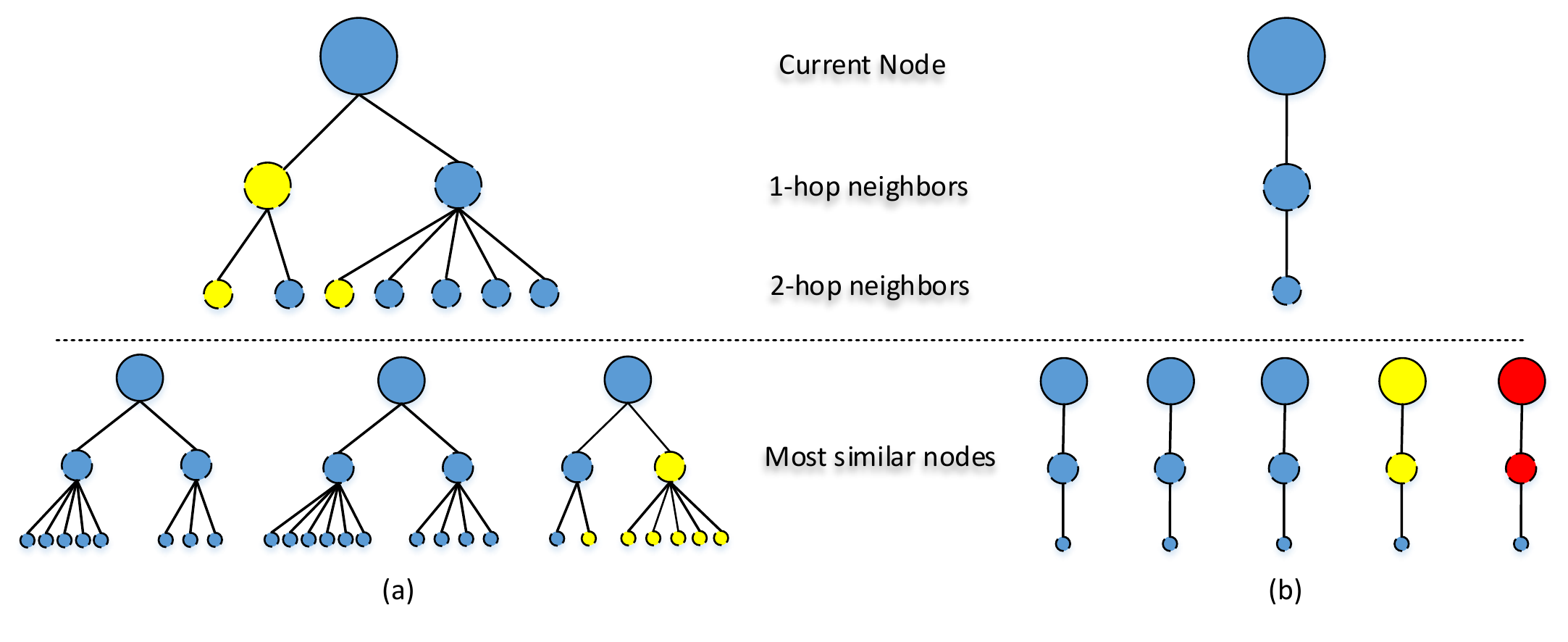}}
\vskip -0.2in
\caption{
Case study for nodes with few neighbors. The top half are the current nodes and their local structures, and the bottom half are the nodes with highest structural
similarities to the current node, as well as their corresponding local structures. Different colors indicate different classes. Both GraphSAGE and
node2vec misclassify the current nodes. In (a) we present the current node with two neighbors, while the detected nodes have similar local and class
information. Same situation occurs in (b).}
\label{fig:case1}
\vskip -0.1in
\end{figure*}

\begin{figure}[tb]
\centerline{\includegraphics[width=0.8\columnwidth]{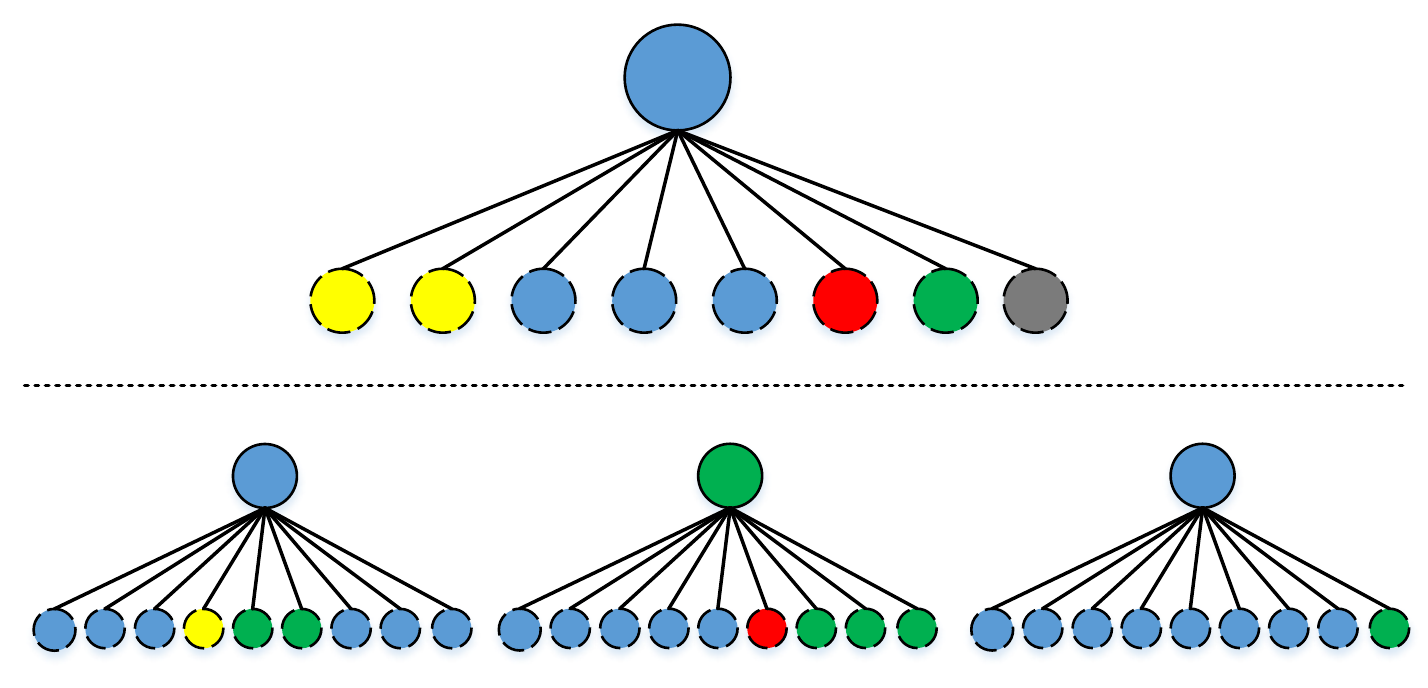}}
\vskip -0.1in
\caption{
Case study for nodes with many neighbors. Following same settings in Figure~\ref{fig:case1}.}
\label{fig:case2}
\vskip -0.1in
\end{figure}

To intuitively analyse the impact of jointly considering local proximity and structural identity on classification performance, we
conduct case studies on the Cora dataset, in the transductive setting. Specifically, we select several representative case from 
nodes that GraphSAGE and node2vec both misclassify. From our observation, the integration of local proximity and structural identity can significantly benefit the classification performance
on two types of nodes: nodes with few or lots of neighbors.

We first illustrate cases of nodes with few neighbors in Figure~\ref{fig:case1}. These nodes are similar to the ordinary users we discussed in Introduction.
(blue nodes in Figure 1), which are also the majority type of nodes in real-world network dataset. Therefore the classification
performance on these nodes has a heavy impact on the overall results. However, normally these nodes have few neighbors, which implies
limited local proximity to utilize, inhibiting the methods that only consider local proximity. In Cora, there are $38.3\%$ nodes
with no more than $2$ neighbors, and this percentage rises to $50.7\%$ among the nodes misclassified by node2vec or GraphSAGE,
indicating it is harder for local proximity methods to deal with ordinary nodes than nodes with more neighbors.

SPINE handles this issue by introducing structural identities. As illustrated in Figure~\ref{fig:case1}, we list the local structure of top $3$
or $5$
nodes with highest structural similarity (computed by line $4$, Algorithm 3) to the current node.
Obviously they have similar local structures, e.g., both have similar numbers of
$1$-hop and $2$-hop neighbors, and more importantly, most of these nodes have the same label with the current node.
SPINE successfully detects these nodes as expected. Thus,
instead of singly leveraging the current node's local information, we are able to
leverage much more structural and local information around these selected nodes. Finally, embeddings with higher quality are learned
and better classification performance are achieved.

Another case of nodes that SPINE can better deal with is those with lots of neighbors but in different classes. These nodes act as
bridges between different academic fields or communities, also know as the structural hole spanners as discussed in Introduction
(red nodes in Figure 1). Therefore, these nodes tend to have many neighbors from different classes, making it hard
to judge their class only depending on the local information, as illustrated in Figure~\ref{fig:case2}.

Again, we alleviate this problem by jointly considering local proximity and structural identity. We also list the $3$ most similar nodes
for each case. From Figure~\ref{fig:case2}, we can observe that the nodes we select not only have similar local structures to the current node,
but also have many neighbors in the same class to the current node at the same time. As a benefit of the extra local structure information,
better performance is achieved in classifying this kind of nodes, comparing to the local proximity methods.

\section{Experiments on Structural Identity}
\subsection{Real World Tasks}
\begin{table}[tb]
\centering
\begin{tabular}{c|c|c}
\hline
Dataset & Pearson (p-value) & Spearman (p-value) \\
\hline
\hline
Citeseer & $0.72$ $(0.0)$ & $0.74$ $(0.0)$ \\
Cora & $0.77$ $(0.0)$ & $0.79$ $(0.0)$ \\
Pubmed & $0.78$ $(0.0)$ & $0.84$ $(0.0)$ \\
\hline
\end{tabular}
\caption{Pearson and Spearman coefficients between structural distance and Euclidean
distance for 
connected node pairs on citation networks.
}
\label{tab:correlation}
\vskip -0.1in
\end{table}

We verify that embeddings learned by SPINE also preserve structural identities in real-world tasks.
We compute the correlation between the structural distance (or similarity) defined in line $4$,
Algorithm 3 and the Euclidean distance in the embedding space for all the connected node pairs.
The values of correlation measured by Pearson and Spearman coefficients are listed in Table~\ref{tab:correlation},
which indicates that there indeed exists a strong correlation between the two distances, validating that SPINE successfully preserves the defined structural similarity in the embedding space.

\subsection{Inductive Setting}
\begin{figure}[tb]
\centerline{\includegraphics[width=0.9\columnwidth]{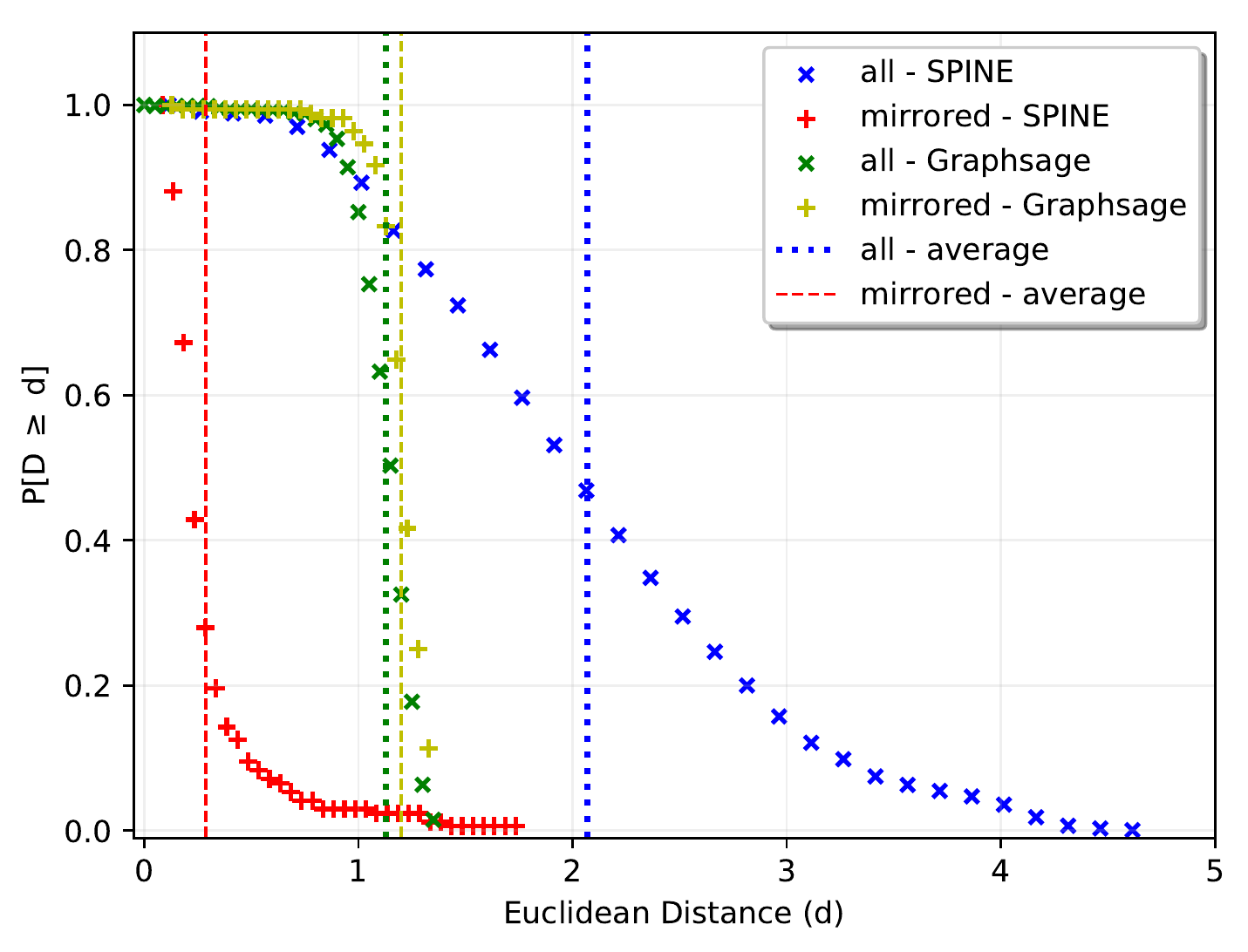}}
\vskip -0.1in
\caption{Euclidean distance distribution between inductively trained embeddings of mirrored node pairs and connected node pairs on the perturbed FB-686 with $s=0.2$.
}
\label{fig:compare_dist_ind}
\vskip -0.1in
\end{figure}

To verify whether SPINE can
capture the structural identity across networks, we generate four new networks $G_{1}$, $G_{2}$, $G_{3}$ and $G_{4}$,
from
the original FB-686 network with $s=0.2$. The combination of $G_{1}$ and $G_{2}$
is considered as training networks, while $G_{3}$ and $G_{4}$ constitute the test data.
After training,
the embeddings of nodes in $G_{3}$ and $G_{4}$ are inferred, on which
we compute the distance distributions between embeddings of 
connected node pairs and mirrored node pairs.
Intuitively, as $G_{1}$ and $G_{2}$ are separated, baselines which only consider local proximities
are not able to capture the structural similarity between networks. 
Results are shown in Figure~\ref{fig:compare_dist_ind}.
Although the structural correlation between training and testing networks is small given $s=0.2$,
the two distance distributions learned by SPINE are still strikingly different,
indicating that SPINE can learn high-level representations of structural identities
from training networks rather than just storing them, which leads to the generalization ability
to identify similar structural identities in unseen networks.
The two distributions learned from Graphsage are practically identical, justifying
our intuition and the necessity of preserving structural identities.

\end{document}